\documentclass[runningheads,a4paper]{llncs}

\usepackage{amssymb}
\usepackage{graphicx}
\usepackage{amsmath}

\usepackage[ruled,vlined]{algorithm2e}
\usepackage{amsmath, amsfonts}

\usepackage{cite}

\usepackage{url}
\usepackage{mathtools}
\urldef{\mailsa}\path|{hung.hoang, bao}@jaist.ac.jp|    

\usepackage[]{algorithm2e}
\usepackage{comment}
\usepackage[hang,small,bf]{caption}
\usepackage[subrefformat=parens]{subcaption}
\usepackage[labelsep=period]{caption}
\SetAlFnt{\small}
\SetKwComment{Comment}{$\triangleright$\ }{}
\let\oldnl\nl
\newcommand\nonl{%
  \renewcommand{\nl}{\let\nl\oldnl}}
\newcommand{\specialcell}[2][l]{%
  \begin{tabular}[#1]{@{}l@{}}#2\end{tabular}}

\begin{document}
\raggedbottom
\mainmatter  

\title{Learning Treatment Regimens \\ from Electronic Medical Records}

\author{Khanh Hung Hoang\and Tu Bao Ho}
\authorrunning{Hoang and Ho.}

\institute{Japan Advanced Institute of Science and Technology,\\
1-1 Asahidai, Nomi, Ishikawa, Japan\\
\mailsa\\
}

\toctitle{Lecture Notes in Computer Science}
\tocauthor{Authors' Instructions}
\maketitle
\begin{abstract}

Appropriate treatment regimens play a vital role in improving patient health status. Although some achievements have been made, few of the recent studies of learning treatment regimens have exploited different kinds of patient information due to the difficulty in adopting heterogeneous data to many data mining methods. Moreover, current studies seem too rigid with fixed intervals of treatment periods corresponding to the varying lengths of hospital stay. To this end, this work proposes a generic data-driven framework which can derive group-treatment regimens from electronic medical records by utilizing a mixed-variate restricted Boltzmann machine and incorporating medical domain knowledge. We conducted experiments on coronary artery disease as a case study. The obtained results show that the framework is promising and capable of assisting physicians in making clinical decisions.

\end{abstract}

\section{Introduction}
The two most important issues in healthcare are disease diagnosis and treatment. While many works have been conducted on the problem of diagnosis prediction, the problem of learning treatment regimens has not yet been extensively studied from the research community. This shortage becomes more serious when hospitals essentially need to make efforts to adopt treatment regimens that best fit their available resources. Additionally, it seems hard to have a fixed care plan for a particular disease due to its high dependency on various patient conditions. As a result, capturing treatment regimens in practice turns out to be meaningful for not only assisting physicians in making right clinical decisions but also helping hospitals manage their resources thoroughly.

In principle, treatment regimens could be learned based on the knowledge-driven approach which requires medical domain or expert knowledge. It can be a piece of information written in the literature or accumulated experience gained by physicians during their career.  While this approach seems to be reliable, taking various domain knowledge into account is costly and not straightforward in reality. In contrast to the knowledge-driven approach, the data-driven approach derives treatment patterns from a large number of observations thanks to the availability of electronic medical records in recent years. Studies followed this approach could be found in \cite{johnston2017retrospective,sun2016data,huang2015mining, merhej2017recommending, song2015penalized}.

Although many interesting results have been achieved, those studies have simply utilized a limited subset of features while many other kinds of data are usually omitted. Such data can be patients' demographics, laboratory test results or clinical notes consisting of signs and symptoms during patients' hospitalization. It is apparent that the more values from those data shared between two patients, the more possibility that the patients are treated with similar regimens.  The lack of considering such valuable information simultaneously in current research could be attributed to the poor-feature data used in their experiments. Moreover, even when the above data is made available, it generally exists in form of numerical, binary, categorical, or text format. Such a heterogeneous data is not ready to use for many data mining methods. Another challenge stems from the fact that treatment regimen is typically defined over periods. Each period is distinct from others at milestones where major changes in a patient's health status happen that lead to a notable adjustment in subsequent prescriptions for the patient. Therefore, given a set of prescription records, identifying suitable treatment periods can considerably affect the learned treatment patterns. 

This work aims to propose a treatment regimen learning framework which addresses both the above challenges. Our framework first divides patients into clusters from which treatment regimens over periods are discovered then. To overcome the challenges of learning from mixed-type data, we employ a mixed-variate restricted Boltzmann machine (MV.RBM) \cite{tran2014mixed}. The advantage of this model is at its robustness in transforming heterogeneous objects to their homogeneous representations. The new latent representations are in the form of hidden binary vectors that could be further processed easily by clustering methods. To tackle the challenge of treatment period identification, we propose an algorithm which can relatively capture significant changes in prescription indications. Moreover, we also suggest another algorithm which derives treatment regimens from each cluster as a regimen tree. The tree can highlight frequently prescribed drugs and infrequently prescribed drugs inside each patient cluster which would be useful for recommending prescribed drugs to patients.

In short, the main contributions of our work are listed as follows. Firstly, we propose a generic framework which can exploit different kinds of relevant patient records. The framework is superior to others in terms of data utilization.  Secondly, we employ both knowledge-driven approach and data-driven approach in our framework. The exploited medical domain knowledge is drug indications and their importance in the treatment for a particular disease. The combination approach used in our framework seems more feasible to deal with the longitudinal property inherent in prescription records. Lastly, we propose a new way to represent treatment regimens flexibly. Frequent drugs are learned from individual level to group level and organized as regimen trees which could be useful for recommending possible regimens to new patients.

\section{Related Work}
This section provides a brief review of studies about the treatment-related learning problem. Notable works can be found in \cite{huang2012mining, huang2015mining, sun2016data}. In \cite{huang2012mining}, the authors developed a process mining method to derive clinical pathway from medical behaviors. Their work, however, mainly focused on learning clinical procedures rather than a detailed treatment.

Inspired by the emergence of electronic medical records, recent studies have exploited prescription records which would provide more useful insights about patient treatment. In \cite{huang2015mining} the authors proposed a probabilistic model that linked patient features and treatment behaviors together to mine treatment patterns. Their model, however, employed many hyperparameters with almost no domain integration. This limitation undermines the model interpretability. Moreover, it was not explicitly described in that work how the chronological order among the learned treatment patterns related. In \cite{sun2016data}, the inspired work of our research, treatment regimens were derived solely from a set of prescription records.  While many typical regimens could be described in an unsupervised mechanism, their prescription-based approach appears to lack of interpretability regarding patient profile and health conditions that lead to the derived regimens. Additionally, although the authors in \cite{sun2016data} attempted to describe the chronological order between regimens with predefined treatment periods, their approach capture little medical domain knowledge as well as seems inflexible in dealing with the varying lengths of hospital stay. Regarding the treatment recommendation task,  \cite{sun2016data} also presented a way to recommend typical treatment regimen for a patient based on demographics and disease severity of patients. This approach, however, seems hard to be applied to new patients whose disease severity may not be recognized at the beginning dates of hospitalization.

\section{Methods}

\begin{figure}[t]
\centering
\includegraphics[height=8cm, width = 9.8cm]{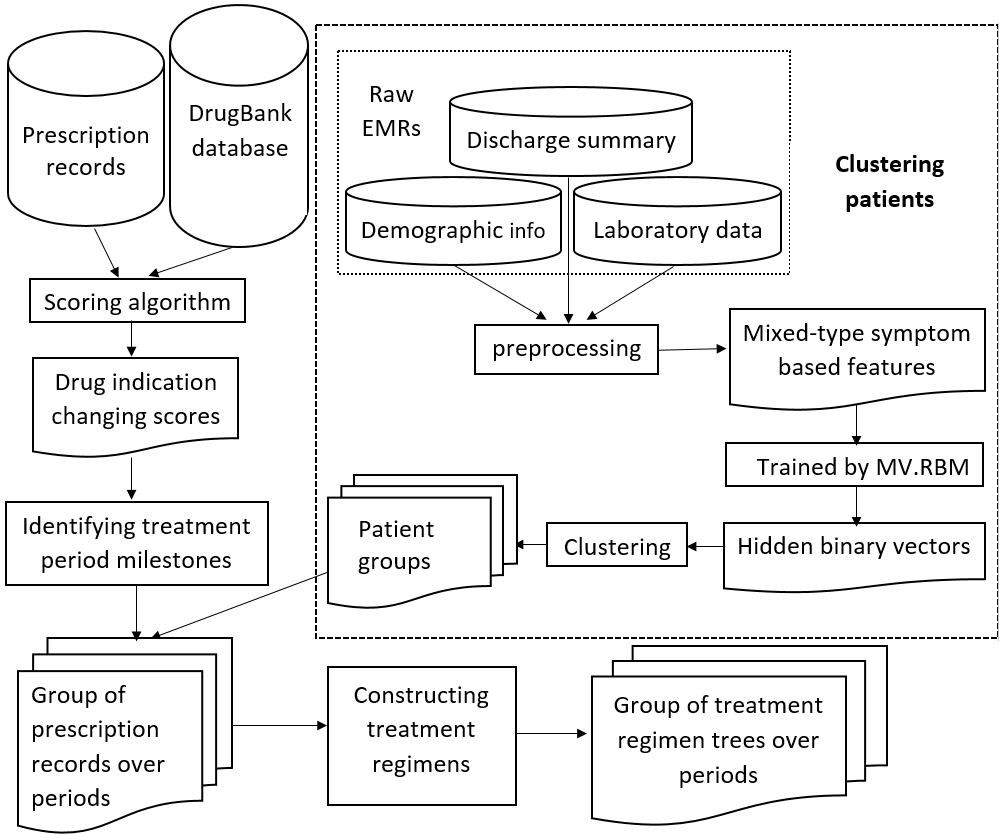}
\caption{An overview of the proposed framework for learning treatment regimens.}
\label{fig:overview}
\end{figure}
 
In this section, we describe our framework of treatment learning problem. This generic framework is designed for a particular disease. Our approach is based on the assumption that a patient cohort may be divided further into groups of more homogeneous patients who share latent characteristics underlying in patient profile or health status. Patients in one group, therefore, are supposed to be treated by similar care plans that share many parts in common.  Fig. \ref{fig:overview} illustrates the framework overview. It consists of two main tasks: clustering a cohort of patients and learning treatment regimens for each resulting cluster.

\subsection{Data Collection and Preprocessing }
Our framework takes medical records of cured patients as trained data. We are interested in the data that characterizes health conditions, for example, demographic information, discharge summary, and laboratory test results. It should be noted that for longitudinal data such as discharge and laboratory indicators, we only collected the records at the early stage of patients since such longitudinal data is usually not fully available for new patients at the time of admission. This solution is based on the intuition that patients who share initial signs, symptoms and laboratory indicators are likely to be treated in the same way. 

After being filtered, patient medical records are encoded as one-hot vectors for categorical data or are normalized to zero-mean unit-variance for numerical data. For discharge summary, only text sections mentioning about the patient history of illness and description about their situation at admission are preferred. We note that segmenting these sections depends on how well-structured discharge summaries were written. In our experiment on MIMIC III database, some clue phrases enabled this solution to become implementable. For simplicity, signs and symptoms mentioned in the segmented text are extracted as new features of the trained patients.  Our framework uses the collection processing engine (CPE) component with  AggregatePlaintextFastUMLSProcessor provided in cTAKES \cite{savova2010mayo}, a well-known tool specifically designed for clinical text processing,  to accomplish this task. It is worth noting that extracted signs and symptoms using this tool links to concepts in the Unified Medical Language System (UMLS) \cite{bodenreider2004unified}, the comprehensive ontology built for the biomedical domain.
\subsection{Data Representation and Patient Clustering}
The encoded data obtained from the previous step contains numerical, binary or categorical values. Such kind of mixed-type data is not ready to fit traditional clustering methods. Therefore, our framework employs MV.RBM, an extension of the restricted Boltzmann machine for data transformation and representation.

MV.RBM is a RBM where visible nodes are not restricted to binary units. Similar to the original RBM, each binary hidden unit in MV.RBM also tries to capture latent aspects in the imhomogeneous visible  units. In other words, MV.RBM could be considered as a model to transform heterogeneous input to homogeneous space. Let  $\boldsymbol{v}=(v_1, v_2,.., v_N)$  denote the set of visible features and $\boldsymbol{h}=(h_1, h_2,.., h_K)$  be the set of hidden units. The energy function of MV.RBM is defined more deliberately to handle  the mixed variate input. \\
$$E(\boldsymbol{v},\boldsymbol{h}) = -(\sum_{i}G_i(v_i)+\sum_{k}b_kh_k+\sum_{ik}H_{ik}(v_i)h_k )$$
where  $\boldsymbol{b}=(b_1, b_2,..,b_N)$ are biases vectors for hidden layer, $G_i(v_i)$ and $H_{ik}(v_i)$ are specified-type functions. By exploiting the conditional independence property within nodes in a layer of bipartite structure, we can get the following factorization equations:
$P(\boldsymbol{v}|\boldsymbol{h}) = \prod_{i=1}^NP(v_i|\boldsymbol{h})$,
 $P(\boldsymbol{h}|\boldsymbol{v}) = \prod_{k=1}^KP(h_k|\boldsymbol{v})$.

The functions $G_i(v_i)$, $H_{ik}(v_i)$ and corresponding $P_i(v_i|\boldsymbol{h})$ for each kind of data are given as follows \cite{tran2014mixed}.

 \begin{table}[h]
 
\begin{center}
\textsf{ \normalsize
\begin{tabular}{|l|c|c|c|} 
               & $G_i(v_i)$ & $H_{ik}(v_i)$   & $P_i(v_i|\boldsymbol{h})$ \\\hline
 Binary        & $a_iv_i$   & $w_{ik}v_i$     & $\frac{\exp(a_iv_i+\sum_kw_{ik}h_kv_i)}{1+\exp(a_i+\sum_kw_{ik}h_k)}$\\[0.2cm]
 Gaussian      & $-v_i^2/2\sigma^2+a_iv_i$    & $w_{ik}v_i$ & $\mathcal{N}(\sigma_i^2(a_i+\sum_kw_{ik}h_k),\,\sigma_i)\ $ \\[0.2cm] 
 Categorical   & $\sum_ma_{im}\delta_m[v_i]$  & $\sum_{m,k}a_{imk}\delta_m[v_i]$& $\frac{\exp(\sum_{m}a_{im}\delta_m[v_i])+\sum_{m,k}w_{imk}\delta_m[v_i]h_k)}{\sum_l\exp(a_{il}+\sum_kw_{ilk}h_k)}$\\  
\end{tabular}
}
\end{center}
\label{tab1}
\end{table}
 
\noindent where $a_{i}$, $a_{im}$ are input bias parameters, $w_{ik}$, $w_{imk}$ are input-hidden weighting parameters. Those with extra subscript $m$ are dedicated for categorical features. 

\begin{figure}
\centering
\includegraphics[height=4cm, width = 9cm]{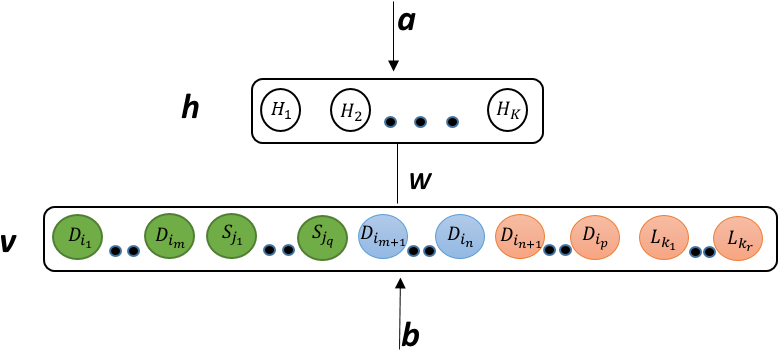}
\caption{A MV.RBM for patient records. The green, blue and orange circles represent for binary, categorical and continuous input units. The circles with labels $D$, $S$, $L$ indicate demographic, signs/symptoms  and laboratory data, respectively.}
\label{fig:rbm}
\end{figure}

In our work, we assume features in the preprocessed data are mutually independent given their latent factors. Fig. \ref{fig:rbm} illustrates our idea to utilize a MV.RBM. We suppose demographic data could receive numerical, binary or categorical values while extracted signs and symptoms are represented as one-hot vectors. Indicator values are assumed to take numerical values. Once the MV.RBM model has been learned, the computable hidden posteriors and hidden states are extracted as transforming features for input $\boldsymbol{v}$. Those latent vectors could be used as input of well-known clustering algorithms. In this concrete work, we utilize the learned binary hidden vectors and select the hierarchical clustering algorithm to divide patients into groups. We use the Hamming distance as similarity measurement for binary vectors and the complete linkage which was reported to give low error rate for symmetric distance measurement \cite{tamasauskas2012evaluation}.

\subsection{Treatment Period Identification}
For each resulting patient cluster obtained from the previous step, prescription records from its patients are collected to derive typical treatment regimens over periods. We represent every drug $dr$ in prescription of patient $p$ as a tuple $dr^{p}= (name, startdate, enddate, dosage)$
that describes drug name, starting date, ending date of usage and its dosage. Let $\Theta^{p} = \{dr^p\}$ be the set of drugs given to the patient, and $T^{p}  = \{dr^p.startdate\}$ be the ordered set of dates the patient $p$ was prescribed. As $|T^{p}|$ varies according to $p$, we propose an algorithm to split each $T^{p}$ into the same number of treatment periods. The idea is for each timestamp in $T^{p}$, we compute an accumulated score that captures the changes in drug indications that have been delivered to the patient so far. We observe the plot of these scores for many patients in the clusters and decide an appropriate number of periods. The splitting dates for each period are the dates with significant changes in their associated scores. 

It is worth noting that in our framework the scoring function takes into account newly prescribed drugs, re-prescribed drugs being stopped using for a while, recently stopped using drugs, or re-prescribed drug with changes in dosage. The aggregate score also gives different weights to those drugs based on their indication. Given a disease $de$ and a set of its common symptoms $Symp^{de}$, we extract from DrugBank database \cite{wishart2007drugbank} the drugs whose indication description  directly mentions about $de$. We name those drugs as main drugs. Drugs with indication mentioned in $Symp$ are also extracted as symptom-healing drugs. Prescribed drugs for the patient $p$ therefore are classified as main drugs, symptom-healing drugs, and unclassified drugs. The weight of each kind of drug is assigned decreasingly according to its importance for the treatment of $de$. We denote $MDB$, $SDB$ as sets of main and symptom-healing drugs which are extracted from DrugBank; $w_{main}$, $w_{symp}$, $w_{unk}$ as the weight for main drugs, symptom-healing drugs, and unclassified drugs, respectively. The detailed algorithm for scoring changes in prescribed drug indications for a patient $p$ is presented in Algorithm~\ref{alg:scoring}. For readability, we remove the superscript $p$ and use Set notations in the pseudocode. 

\begin{algorithm}[H]
{\small
 \KwData{$\Theta$, $T$, $MDB$, $SDB$}
 \KwResult{return $scores$ as a list of accumulated scores}
 Initialize  $U$ as an empty set \Comment*[r]{set of recently delivered drugs}
 Initialize $scores$ as an empty list \;
 $aScore := 0$   \Comment*[r]{the accumulated score}
 \For{ each $d \in T$}{
  $D := \{dr\:| \: \forall dr \in \Theta \land  dr.startdate == d \}$ \Comment*[r]{delivered drugs on date d}
  $N:= \{dr\:| \: \forall dr \in D \land  dr.name \notin U.name \}$ \Comment*[r]{ newly  delivered drugs}
  $DC := \{dr \: | \: \forall dr \in D,  \exists dr' \in U\: \text{such that} \: dr.name == dr'.name \land  dr.dosage <> dr'.dosage \}$ \Comment*[r]{dosage changed drugs}
  $S:= \{dr \: | \: \forall dr \in U \land dr.name \notin D.name \land dr.enddate < d \}$ \Comment*[r]{ recently stopped using drugs}
  \For{ each $d' \in U$}{
  		\If{$\exists d'' \in D \: \text{such that} \: d'.name==d''.name$}{
  			$d' := d''$ \Comment*[r]{update U with redelivered drugs}
   }
  }
   $U := (U \setminus S)  \cup N $ \Comment*[r]{update U with newly delivered drugs}
   
   $CD := N \cup DC \cup S$ \Comment*[r]{considering drugs for calculating scores}
   $CMD := CD.name \cap MDB$\Comment*[r]{considering main drugs}
   $CSD := CD.name \cap SDB$\Comment*[r]{considering symptom-healing drugs}
   $UD := CD.name \setminus (CMD \cup CSD)$\Comment*[r]{unclassified drugs} 
   $aScore = aScore + |CMD| \times w_{main} + |CSD| \times w_{symp} + |UD| \times w_{unk}$\;
   Add $aScore$ to  $scores$
 }
 }
 \caption{Scoring prescription records}
 \label{alg:scoring}
\end{algorithm}
\subsection{Learning Group Treatment Regimens}
The previous section has demonstrated our domain integrated algorithm which allows prescription recorded to be divided into periods based on the associated scores which reflect the change in the indication of prescribed drugs. In this section, we describe how a treatment regimen over a period of a given patient cluster is derived. We relax the chronological order of delivered drugs in a period and restrict the element of constructed treatment regimens to drug names only. Other information such as dosage, route, is assumed to be decided by the physicians. 

The learned regimens were organized in a tree form. Starting from the root, we assign the most frequently prescribed drug $d$ to its left child node and extract prescribed drugs excluding $d$ of the patients who were treated by  $d$. The drug assignment for next right child nodes will follow the similar approach applied on prescribed drugs of those patients who were not treated by left-hand side nodes in the same level. We recursively perform this procedure on internal nodes. To avoid learning too complicated details of the derived tree, we only perform the procedure until a certain level of the tree or when the number of patients treated by the most frequent drug for the parent node is still greater than a threshold. Algorithm~\ref{alg:tree} presents our ideas to construct the treatment regimen tree for a particular group of patients in a period. 

\begin{algorithm}
\SetKwFunction{regimenTree}{\textsc{regimen}-Tree}
\Indm\nonl\regimenTree{$depth, prescData,  parent, traces$}\\
\Indp
	\If{$prescData$ is empty or $depth==maxDepth$}{\KwRet{}}
	$d$ := most frequent drugs from $prescData$ \;
	$nPatients$ := number of patients who  were treated by $d$\;
	$traces[parent, d]=\text{``}\nwarrow\hspace{-3pt}\text{''}$\;
	$cNodePresc$ := prescribed records excluding $d$ of patients treated by $d$ \;
    $rNodePresc$ := prescribed records of patients who were not treated by $d$ \;
   \uIf{$nPatients<threshold$ }{
   \regimenTree{$depth, rNodePresc, parent, traces$}\;
   
  }
  
  \Else{
    \regimenTree{$depth + 1, cNodePresc, d, traces$}\;
    \regimenTree{$depth, rNodePresc, parent, traces$}\;
  }
  \KwRet{
    }\;
    \caption{Procedure for the construction of a treatment regimen tree}
   \label{alg:tree}
\end{algorithm}
\section{Experimental Evaluation}
This section presents our experimental evaluation of the proposed framework for deriving typical treatment regimens from electronic medical records. The obtained results of the clustering analysis, treatment period identification and learned treatment regimen trees are also given and analyzed.  Lastly, we propose a method to evaluate the efficacy of the derived treatment regimen trees in recommending prescribed drugs for new patients.
\subsection{Experimental Design}
Our experimental evaluation was performed on MIMIC III, a freely accessible critical care database \cite{johnson2016mimic}.  We considered the treatment regimen of patients who were diagnosed with coronary artery disease as a case study. Since a patient could be diagnosed with multiple ICD codes, to ensure the homogeneity of our patient cohort, we only selected those whose primary ICD is coronary artery disease and comorbidity scores are zero for other disease groups. In addition, patients who were prescribed fewer than three times were also excluded from the experimental evaluation. The number of extracted patients is 707 of which we randomly selected  687 patients for training and left 20 patients for testing the efficacy of the learned treatment regimens. We followed the approach described in the Data Representation section to preprocess raw data. A summary of preprocessed data with illustration features is given in Table \ref{tab2}

 \begin{table}
 
\begin{center}
\textsf{ \small
\begin{tabular}{|l|l|} 
 Kind of data (no.features)  & Sample features (data type)  \\\hline
 Demographic info (11)        & \specialcell{age (numerical), gender (binary)\\admission type (categorical)}  \\  \hline
 Laboratory data (175)        & \specialcell{arterial blood pressure (numerical)\\ atrial pacemaker(numerical)}  \\  \hline
 Signs and symptoms data (1466)        & \specialcell{abdominal discomfort (binary)\\ ability to climb (binary)\\able to sleep (binary)} \\ 
\end{tabular}
}
\end{center}

\caption{ A short summary of features in the dataset}
\label{tab2}
\end{table}

We fit preprocessed data as input for  MV.RBM with 200 hidden units since the trained error did not decrease significantly with a larger number of hidden units. The learned binary hidden states were then extracted as representation features for the subsequent clustering task. We employed hierarchical clustering with parameters are described in the previous section.  For the task of treatment period identification, we extracted main drugs and symptom-healing drugs from DrugBank database. The referred typical symptoms of coronary artery disease in the literature are ``heart attack'',``shortness of breath'' and ``chest pain''. We assigned the weight of main drugs, symptom healing drugs, and unclassified drugs to 1, 0.5 and 0.1, respectively. The threshold of cutting node in treatment learning algorithm was set to 10 patients. In our experiment, we derived regimens until a certain level of the tree. The depth parameter was set to 4. 

\subsection{Results}
\begin{figure}[b]
\centering
\includegraphics[height=3.5cm, width = 9cm]{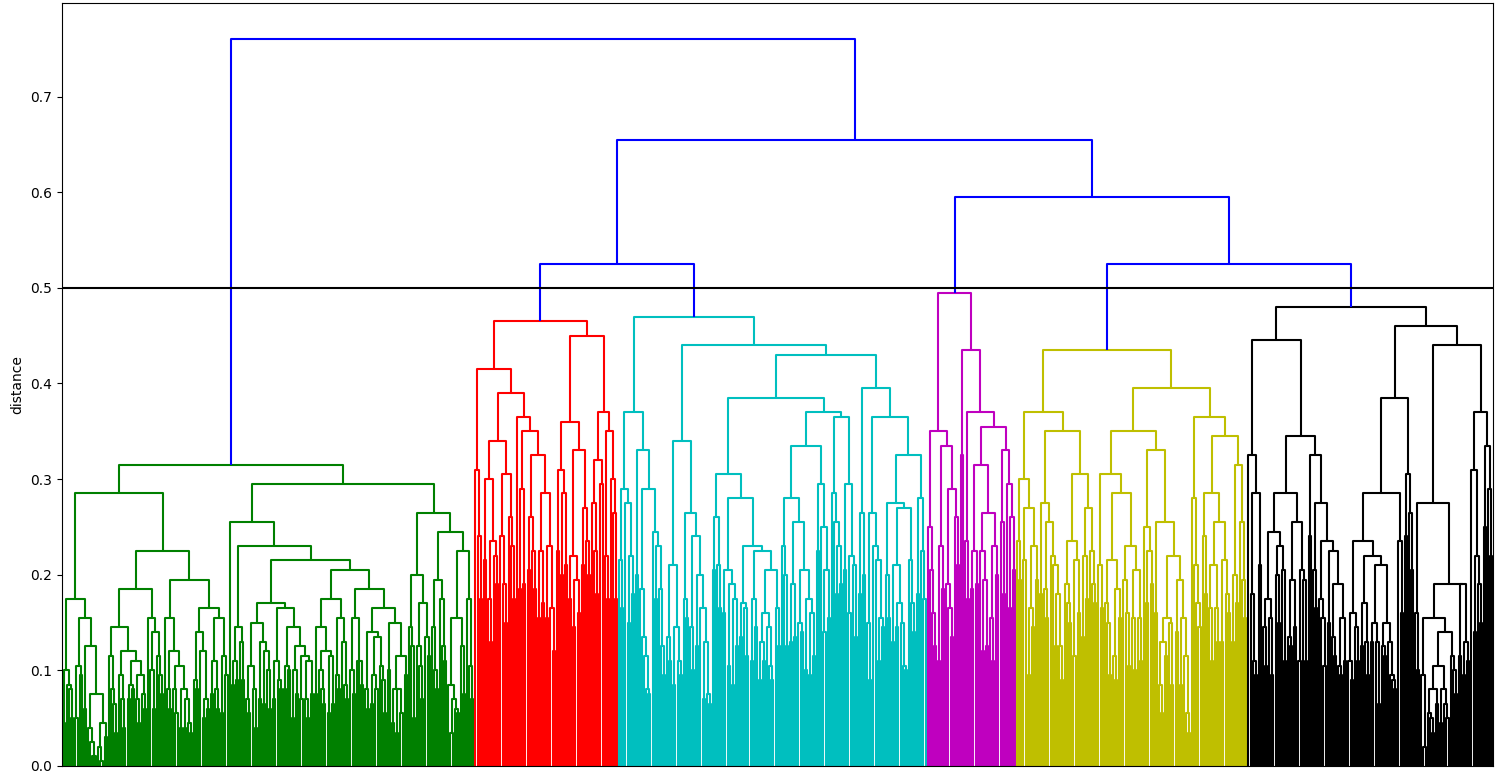}
\caption{Dendrogram of hierarchical cluster analysis 
}
\label{fig:clustering}
\end{figure}

Fig. \ref{fig:clustering} describes a dendrogram of clustering results. It is noted that the trained patients themselves are homogeneous subjects in terms of diagnostic perspective. Thus, we preferred a relatively small number of clusters. Based on the visualization, we decided to group the trained patients into six clusters. The size of each cluster is 198, 69, 148, 43, 111 and 118 patients, respectively.

\begin{figure}[t]
\centering
\includegraphics[height=3cm, width = 9cm]{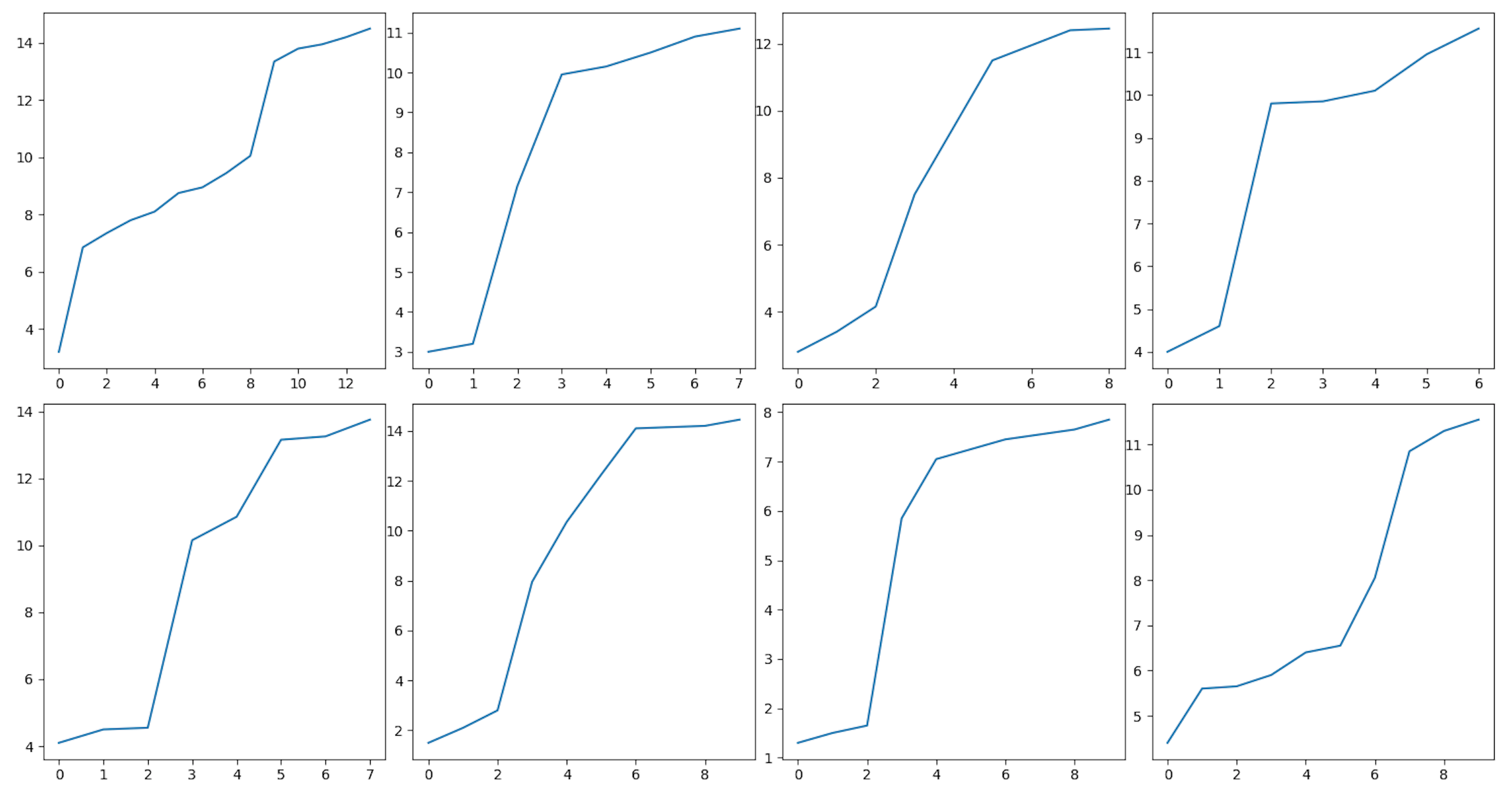}
\caption{Sample line charts of accumulative scores of randomly taken patients. For each line chart, the vertical axis represents the accumulative scores, the horizontal axis represents the timestamps when a patient was prescribed.
}
\label{fig:scoring}
\end{figure}

\begin{figure}
\centering
\includegraphics[height=4cm, width = 12cm]{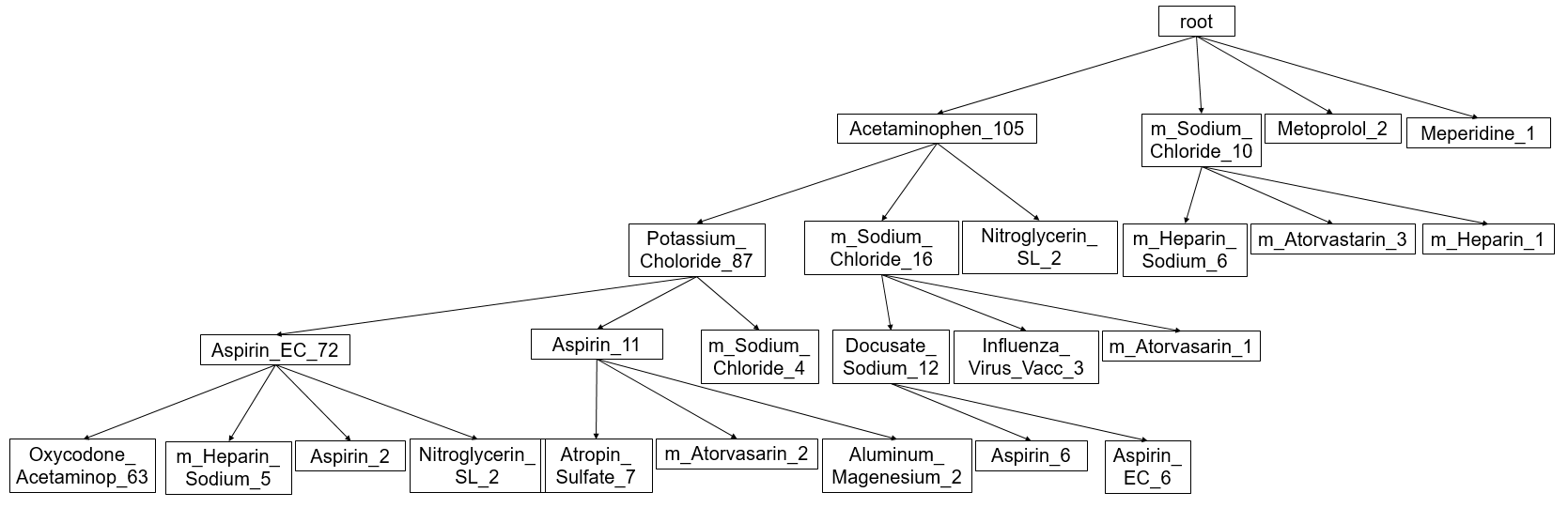}
\caption{Sample of learned regimen tree. The prefix ``m'' denotes for main drugs while the last integer denote the number of prescribed patients.
}
\label{fig:regimentree}
\end{figure}

Fig. \ref{fig:scoring}  presents a few randomly taken line charts of accumulative scores for eight patients. Interestingly, most of the plots follow similar patterns. There is a  slight increase in scores at the beginning and the end of every treatment compared to the significant change at the center interval. Therefore, we decided to divide prescription records of trained patients into three periods. Fig. \ref{fig:regimentree} illustrates an example of constructed regimen trees. Given a path in the tree, we note that the order of the nodes in this path should be understood as frequency order of drug use rather than chronological order of prescription time. It can be seen that the visualization can provide hint-drugs probably delivered together with a given drug. Therefore, physicians can use the learned trees as a checklist to decide which drugs are likely and unlikely to be prescribed.
\subsection{Evaluation}
We evaluate the efficacy of learned regimen trees in recommending prescribed drugs to new patients. It should be noted that patient records of the testing set are represented by the trained MV.RBM. We consider the patients in each resulting clusters as labeled data and assign the cluster index for test patients based on their nearest neighbors. Given a new patient $p$, let $p'$ be his/her nearest neighbor which has been assigned to cluster $c_i$. The recommended drugs should be given to $p$ in a particular period are drugs on the path of regimen tree of $c_i$ in the same period such that $p'$ was prescribed with each drug on that path.

Let $T$, $nP$,  $\hat{D}_{p}^{t_j}$, and $D_{p}^{t_j}$ denote the test set, the number of periods, the recommended path, i.e., the set of recommended drugs for $p$ over period $t_j$, and  the set of prescribed drugs for $p$ in that period, respectively.  We propose two measures to evaluate the efficacy of learned regimens for the prescription recommendation task. These measures reflect how likely  $\hat{D}_{p}^{t_j}$  is a subset of  $D_{p}^{t_j}$. $\hat{D}_{p}^{t_j}$  is said ``correct'' if it is a non empty subset of $D_{p}^{t_j}$. In case  $\hat{D}_{p}^{t_j}$  has non empty intersection with  $D_{p}^{t_j}$ but not its subset, we say the set $\hat{D}_{p}^{t_j}$ is ``approximately correct''. We denote $m_{cor}$ as the percentage of recommended paths which are ``correct'' and $m_{app}$ as the percentage of  recommended drugs actually prescribed in both ``correct'' and ``approximately correct'' paths. Let $I_A(B)$ define the indicator function which return $1$ if  $B \subset A$ or  $0$ otherwise.  We have:
$$\mbox{\small $m_{cor}=\frac{1}{|T|\times nP} \sum_{p \in T}\sum_{j=1}^{nP}I_{D_{p}^{t_j}}(\hat{D}_{p}^{t_j})$}; \text{ }\mbox{\small $m_{app}=\frac{1}{|T|\times nP} \sum_{p \in T}\sum_{j=1}^{nP}\frac{|\hat{D}_{p}^{t_j}\cap D_{p}^{t_j}|}{|\hat{D}_{p}^{t_j}|}$}$$ 

We repeated our experiment 10 times for different training and testing sets. The obtained values of $\bar m_{cor}$ and $\bar m_{app}$ are $0.527$ and $0.729$, respectively. Although the obtained values of $\bar m_{cor}$ should be further improved, to some extent, these measures show the efficacy of the regimen trees derived from our proposed framework. 
\section{Discussion}
Comparing to related works in the literature, our work obtained more interesting results in terms of domain exploitation and knowledge representation. Rather than defining a similarity metric by a frequency-based approach for complex objects \cite{sun2016data}, we tracked the change of drug indication in prescribed drugs as a hint to discover treatment periods. It can be seen that the idea fits our natural thinking on detecting patients' treatment periods given their prescription records. The common pattern found in  Fig   \ref{fig:scoring} has reconfirmed the rationality of our proposed domain-based algorithm. Moreover, representing the learned regimens in form of trees not only fully reflects the usage-frequency of drugs but also allows doctors to quickly recognize groups of frequently and infrequently prescribed drugs in each patient sub-cohort. Therefore, in terms of knowledge representation, it could be said that our work is superior to  \cite{huang2015mining, sun2016data} where the authors simply organized treatment patterns in flat form.

There are several reasons to explain the primitive results of our initial study on the task of treatment recommendation. Firstly, it is worth noting that we addressed the problem of treatment recommendation on MIMIC III, a practical and very challenging dataset. Even if it has been simplified to recommend in total up to 12 among many prescribed drugs for every patient, the problem is still not trivial as there are hundreds of different drugs given in the prescription records. Additionally, while our evaluation metrics directly assess whether the recommended drugs are prescribed to new patients, it is not clearly described in other studies how well the recommended treatments match the actual prescribed drugs.  We leave the task of improving our prediction accuracy with a more deliberated framework for the future work. 
\section{Conclusion}
In this paper, we have presented a generic framework to derive treatment regimens from electronic medical records. The proposed framework is novel in terms of data utilization, domain incorporation, and regimen representation. The experimental evaluation has shown the efficacy of learned treatments for the task of prescription recommendation. Although further improvement should be made such as data cleaning and normalizing for clinical features, this study is a pioneering work which encourages researchers to exploit medical domain knowledge and address the treatment learning problem more thoroughly.
\subsection*{Acknowledgement} This work is partially sponsored by Asian Office of Aerospace R\& D under agreement number FA2386-17-1-4094 and Vietnam National University at Ho Chi Minh City under the grant number B2015-42-02. We wish to thank Tu Dinh Nguyen for providing the implementation of the MV.RBM model. 
\bibliographystyle{splncs04}
\bibliography{654.bib}
\end{document}